\documentclass{appolb}
\usepackage{epsfig}

\usepackage{amsmath}

\newcommand{\PL}[1]{{ Phys.\ Lett.\ } {\bf  #1}}

\newcommand{\PRL}[1]{{ Phys.\ Rev.\ Lett.\ } {\bf  #1}}

\newcommand{\lsim}{\raise.3ex\hbox{$<$\kern-.75em\lower1ex\hbox{$\sim$}}}
\newcommand{\ima}{{\mbox{Im}\,}}

\begin{document}
\title{Light scalar mesons: comments on their behavior in the $1/N_c$ expansion near $N_c=3$
versus the $N_c\rightarrow\infty$ limit%
}
\author{ J. R. Pel\'aez\thanks{Invited talk at "`Excited QCD"', 8-14 February. Zakopane, Poland.} and G. R\'{\i}os
\address{Dept. de F\'{\i}sica Te\'orica II. Universidad Complutense. 28040 Madrid. Spain}
}
\maketitle
\begin{abstract}
We briefly review how light meson resonances are described within one and two-loop
Unitarized Chiral Perturbation Theory amplitudes and how, close to $N_c=3$, light vectors 
follow the $N_c$ behavior of $q\bar q$ mesons whereas light scalars do not.
This supports the hypothesis that the lightest scalar is not predominantly a $q\bar q$ meson,
although a subdominant $q\bar q$ component is suggested around 1 GeV at somewhat larger $N_c$.  In contrast, when $N_c$ is very far from 3, like in the $N_c\rightarrow\infty$ limit, we explain again in detail why unitarization is not, a priori, reliable nor robust 
and why this limit should not be used to drag any conclusions about the dominant nature of physical light
scalar mesons.
\end{abstract}
  
\vspace{-.2cm}
\section{Introduction}
The $1/N_c$ expansion \cite{'tHooft:1973jz}
is the only analytic expansion of QCD in the whole energy region, 
that provides a definition of $\bar qq$ bound states, 
whose masses and widths behave as $O(1)$ and $O(1/N_c)$, respectively.
Light mesons are also described within Chiral Perturbation Theory (ChPT)\cite{chpt1}, which
is the QCD low energy effective theory, built as 
the most general effective lagrangian compatible with 
all QCD symmetries, involving
the pseudo Nambu-Goldstone Bosons of the QCD 
spontaneous chiral symmetry breaking. Meson-meson scattering amplitudes become
an expansion in momenta and masses, generically denoted $p$, over a scale 
$\Lambda_\chi\sim 4\pi f_\pi\simeq 1$ GeV. At each order, the ChPT Lagrangian contains
\emph{all terms} compatible with QCD symmetries,
multiplied by Low Energy Constants (LECs), that encode the QCD dynamics and renormalize
divergences order by order. 

The correct QCD leading order $1/N_c$ behavior of $f_\pi$, the
pseudo Nambu-Goldstone boson masses and the LECs, is well known and ChPT amplitudes 
have no cutoffs or subtraction constants
where spurious $N_c$ dependences could hide.
Note that, in order to apply the $1/N_c$ expansion, 
the renormalization scale $\mu$
has been chosen between $\mu=$ 0.5 and 1 GeV, following \cite{chpt1}.
(Also, in Fig. \ref{ncSU3} we show that outside this band 
the generated vector mesons will start deviating from their well established $\bar qq$ behavior).

Resonances are not present in the ChPT lagrangian but can be described using ChPT as input in
a dispersion relation
\cite{Dobado:1996ps}. 
The main idea is that 
partial waves,
$t$, of definite isospin and angular momentum satisfy
an elastic unitarity condition: 
$\ima 1/t=-\sigma,$
while the ChPT  expansion
$t\simeq t_2+t_4+\cdots$, $t_k=O(p^k)$, 
satisfies it only
perturbatively: 
$\ima t_2=0, \ima t_4=\sigma\vert t_2\vert^2,\cdots$.

Since $G=t_2^2/t$ has a right cut ($RC$)
a left cut ($LC$), and possible
pole contributions ($PC$), we can write a dispersion relation as follows
\begin{equation}
  \label{disp1/t}
  G(s)=G(0)+G'(0)s+\tfrac{1}{2}G''(0)s^2+
  \frac{s^3}{\pi}\int_{RC}ds'\frac{\ima G(s')}{s'^3(s'-s)}+
  LC(G)+PC.
\end{equation}
In the elastic approximation, unitarity allows us to evaluate \emph{exactly} 
$\ima G=-\sigma t_2^2=-\ima t_4$ on the $RC$. The subtraction
constants can be approximated with ChPT since they
involve amplitudes evaluated at $s=0$, 
$G(0)\simeq t_2(0)-t_4(0),\cdots$.
These three subtractions imply that $LC$ is dominated by its
low energy part, and well estimated by ChPT as
$LC(G)\simeq LC(-t_4)$. 
$PC$ counts as $O(p^6)$ and only gives sizable contributions much below
threshold in scalar waves \cite{GomezNicola:2007qj}, thus we neglect it here for simplicity.
All in all, one finds the IAM formula \cite{Dobado:1996ps}:
\begin{equation}
  \label{eq:IAM}
  t(s)\simeq \frac{t_2^2(s)}{t_2(s)-t_4(s)}\,.
\end{equation}
Remarkably, this simple equation ensures elastic unitarity, matches ChPT
at low energies, describes fairly well data up to somewhat less than 1 GeV, and generates
the $\rho$, $K^ *$, $\sigma$ and $\kappa$ resonances
as poles on the
second Riemann sheet, with ChPT parameters rather similar to
those from standard ChPT.
The IAM can be easily
extended to higher orders or -- without a dispersive justification yet -- generalized within a coupled channel formalism 
\cite{GomezNicola:2001as,Oller:1997ng}, 
generating also the $a_0(980)$, $f_0(980)$ and the octet $\phi$.

By scaling with $N_c$ the ChPT parameters in the IAM, we can determine
the $N_c$ dependence of the resonances masses and widths \cite{Pelaez:2003dy,Pelaez:2006nj},
defined from the pole position as $\sqrt{s_{pole}}=M-i\Gamma$,
and compare it with the $\bar qq$ scaling to determine if
the resonance is predominantly of a $\bar qq$ nature.

However, {\it a priori}, one should be careful {\it not to take $N_c$ too large, and in particular
to avoid the $N_c\to\infty$ limit, because it is a weakly interacting limit}. As shown above, the IAM relies
on the fact that the exact elastic $RC$ contribution dominates the
dispersion relation.
Since the IAM describes the data and the resonances, within, say 10 to 20\% errors, this means that at $N_c=3$ the other contributions are not
approximated badly.  But meson loops, responsible for the $RC$, scale as $3/N_c$ whereas the inaccuracies due to the approximations scale partly as $O(1)$.
Thus, we can estimate that those 10 to 20\% errors at $N_c=3$ may become 100\% errors at, say $N_c\sim30$ or $N_c\sim15$, respectively.
Hence we have never shown results \cite{Pelaez:2003dy,Pelaez:2006nj} beyond $N_c=30$, and even beyond $N_c\sim15$ they should be interpreted with care.

Of course, there could be special cases in which the IAM could still work for very large $N_c$,
as it is has been shown for the vector channel for QCD \cite{Nieves:2009ez}. 
But that is not the case for the scalar channel, which, 
if used for too large $N_c$ may lead to inconsistencies \cite{Nieves:2009ez}
for some values of the LECS.

\vspace{-.4cm}
\section{$N_c$ scaling of resonances}
\begin{figure}[t]
  \centering
  \vbox{
    \hbox{
      \includegraphics[scale=.43]{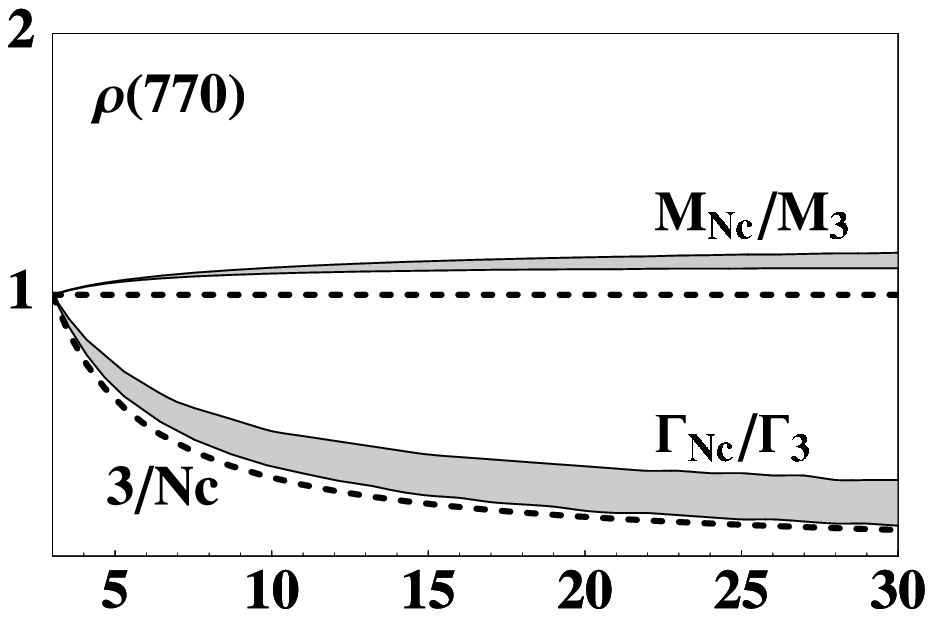}
      \includegraphics[scale=.43]{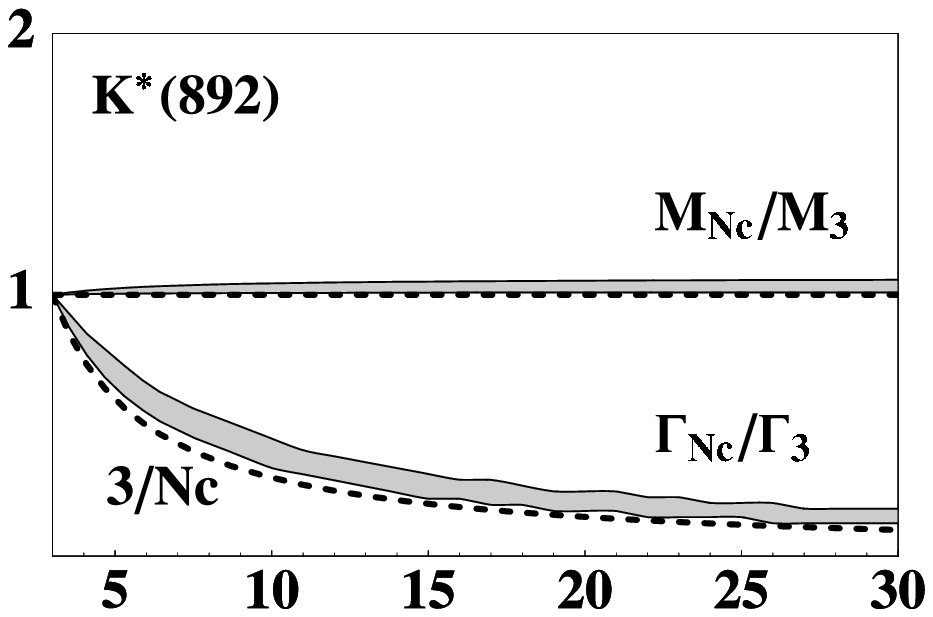}
      \includegraphics[scale=.43]{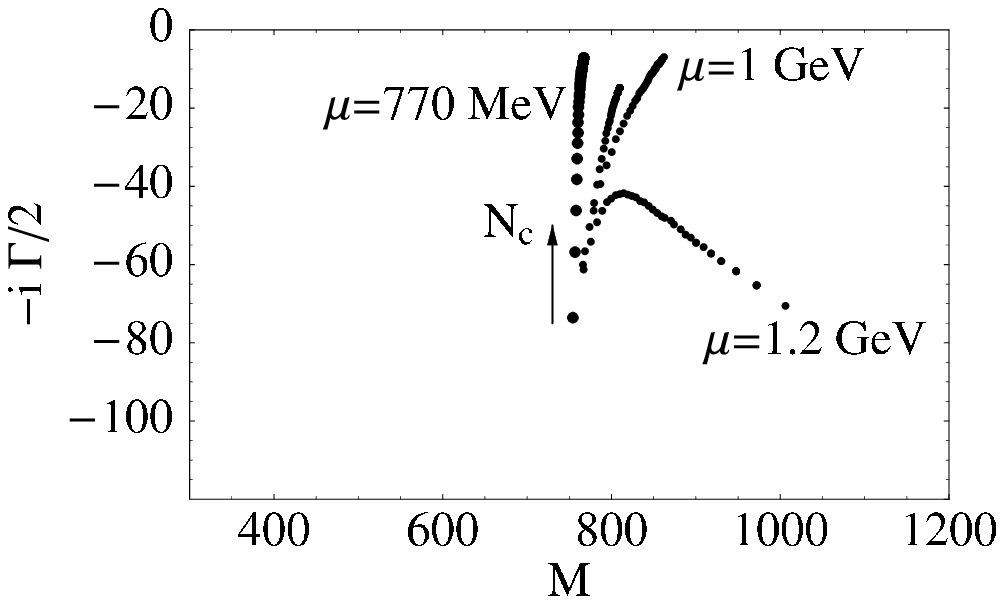}
    }
    \hbox{
      \includegraphics[scale=.43]{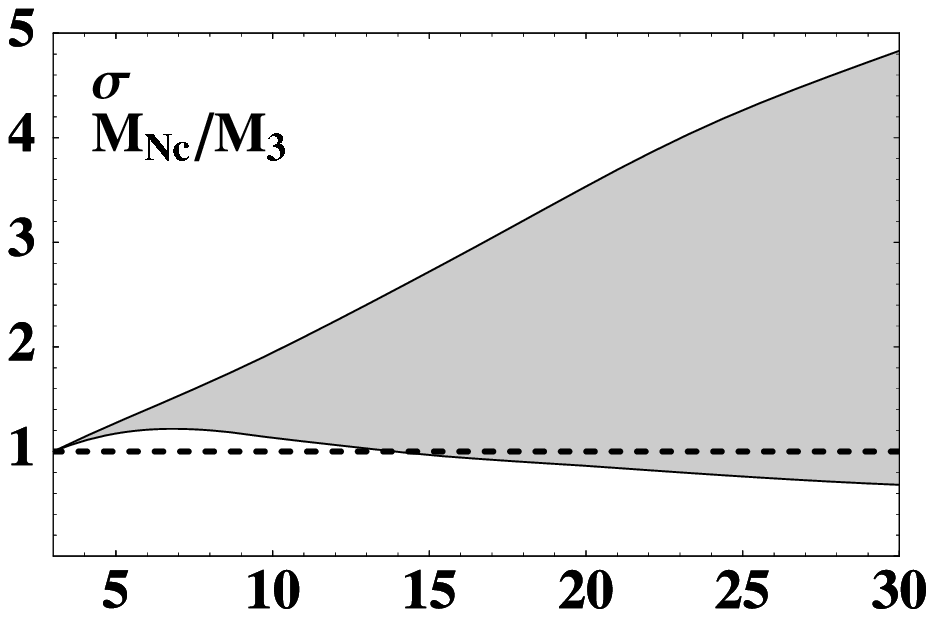}
      \includegraphics[scale=.43]{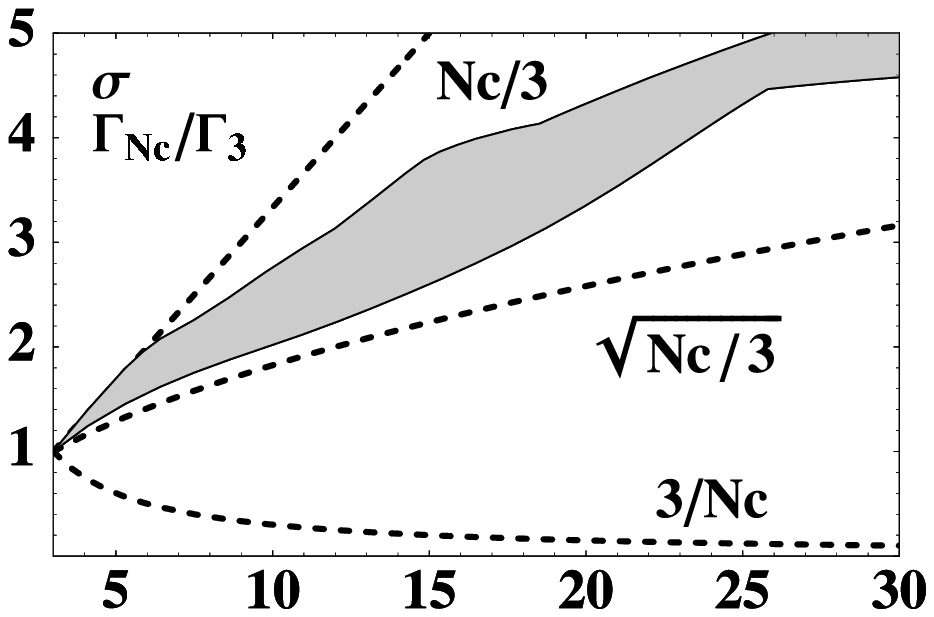}
      \includegraphics[scale=.43]{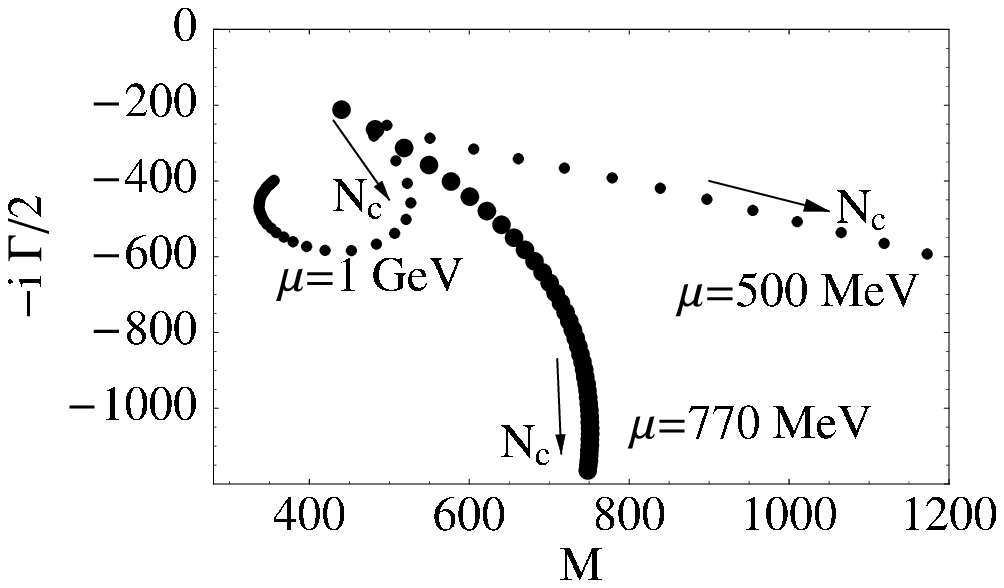}
    }
  }
  \caption{{\bf Top:} $N_c$ behavior of the $\rho$ and $K^*$
  mass and width (left and center). Right: Different $\rho$ pole
  trajectories for different values of $\mu$, note that for
  $\mu=1.2$ GeV the $\rho$ pole goes away the real axis.
  {\bf Bottom:} $N_c$ behavior of the $\sigma$ mass and width
  (left and center). Right: Different $\sigma$ 
  pole trajectories for different $\mu$ values.}
  \label{ncSU3}
\end{figure}

The $N_c$ scaling of IAM resonances 
was studied to one-loop in coupled channels in \cite{Pelaez:2003dy}
and to two-loops in the elastic case in \cite{Pelaez:2006nj}.
Thus, Fig.\ref{ncSU3} shows the
behavior of the $\rho$, $K^*$ and $\sigma$ masses and widths
found in  \cite{Pelaez:2003dy}. The $\rho$ and
$K^*$ neatly follow the expected behavior for 
a $\bar qq$ state: $M\sim 1$, $\Gamma\sim 1/N_c$.
The bands cover the uncertainty
in $\mu\sim 0.5-1$ GeV
 where to scale the LECs with $N_c$.
Note also in Fig.1(Top-right) that, 
for that set of LECS, {\it outside this $\mu$ range}
the $\rho$ meson starts deviating from a 
a $\bar qq$ behavior. Something similar occurs to the $K^*(892)$.
Consequently, we cannot
apply the $N_c$ scaling at an arbitrary $\mu$ value,
if the well established $\rho$ and $K^*$ $\bar qq$ nature is to be reproduced.

In contrast, the $\sigma$ shows a 
different behavior from that of a pure $\bar qq$:
\emph{near $N_c$=3} both its mass and width
grow with $N_c$, i.e. its pole moves 
away from the real axis. 
Of course,  far from $N_c=3$, and
for some choices of LECs and $\mu$,
the sigma pole might turn back to the real axis \cite{Pelaez:2006nj,Nieves:2009ez,Sun:2004de}, 
as seen in Fig.1 (Bottom-right). 
But, as commented above, the IAM is less reliable for large $N_c$,
and even if we trust this behavior it only suggests that
there might be a subdominant $\bar q q$ component \cite{Pelaez:2006nj}.
In addition, we have to make sure that the LECs used fit data
and reproduce the $\bar qq$ behavior for the vectors.

Since loop terms are important in determining the scalar pole position,
but are $1/N_c$ suppressed compared to tree level terms with LECs,
it is relevant to check the  $O(p^4)$ results
with an $O(p^6)$ IAM calculation. This was done
 within $SU(2)$ ChPT in \cite{Pelaez:2006nj}. 
We defined a $\chi^2$-like function to measure how close 
 a resonance is from a $\bar qq$ $N_c$  behavior.
First, we used that $\chi^2$-like function at $O(p^4)$ to show
 that it is not possible for the $\sigma$ to behave predominantly as a $\bar q q$
while describing simultaneously the data
and the $\rho$ $\bar qq$ behavior, thus
{\it confirming the robustness of the conclusions for $N_c$ close to 3}. 
Next, we obtained a $O(p^6)$ data fit -- where
the $\rho$ $\bar qq$ behavior was imposed -- whose $N_c$ behavior for the
$\rho$ and $\sigma$ mass and width is shown in Fig.2. Note that
both $M_\sigma$ and $\Gamma_\sigma$ grow with $N_c$, near $N_c=3$
confirming the $O(p^4)$ result of a non $\bar qq$ dominant component.
However, as $N_c$ grows further, between $N_c\sim8$ and $N_c\sim 15$, where we still
trust the IAM results, $M_\sigma$
becomes constant and $\Gamma_\sigma$ starts decreasing. 
This may hint to a \emph{subdominant $\bar qq$ component},
arising as loop diagrams become suppressed as $N_c$ grows.
Finally, we checked how big this $\sigma$ $\bar qq$ component
can be made, by forcing the $\sigma$ to behave 
as a $\bar qq$ using the above mentioned $\chi^2$-like measure.
We found that in the best case, this subdominant $\bar qq$
component could become dominant around $N_c>6$, at best, but
always with an $N_c\to\infty$ mass  above roughly 1 GeV instead of its physical $\sim 450$ MeV value.

\begin{figure}[t]
  \centering
  \hbox{
    \includegraphics[angle=-90,scale=.34]{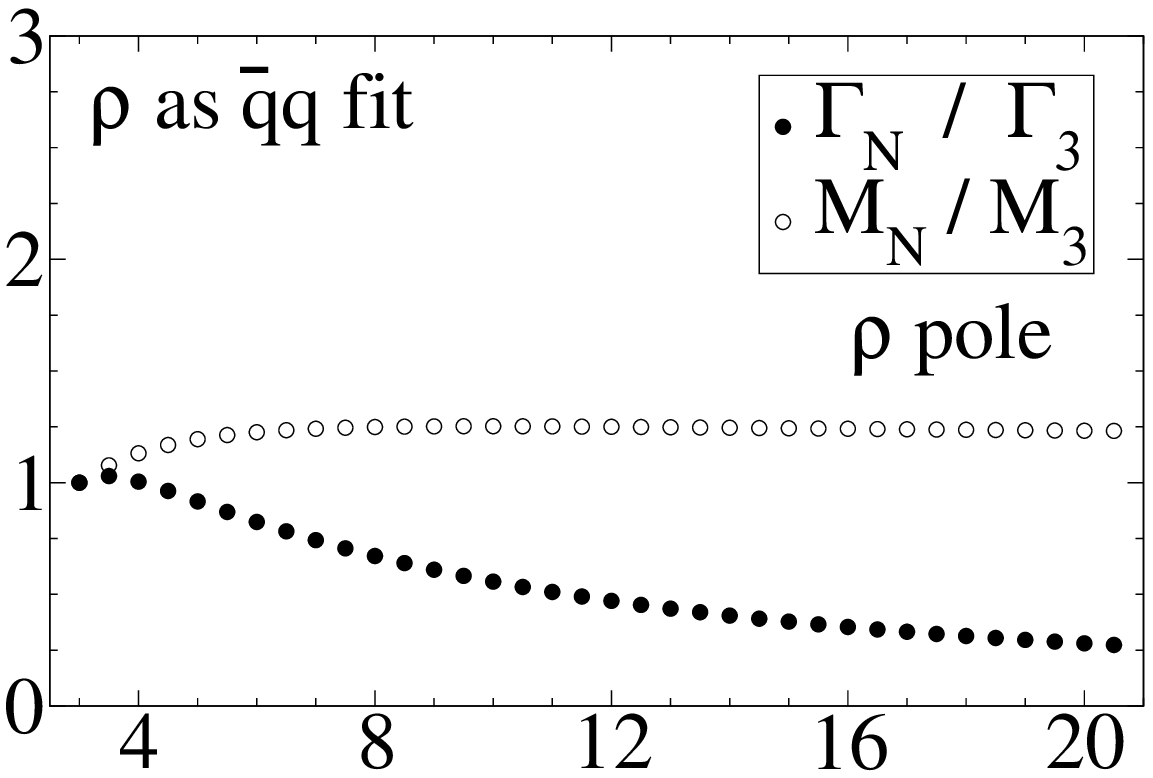}
    \includegraphics[angle=-90,scale=.34]{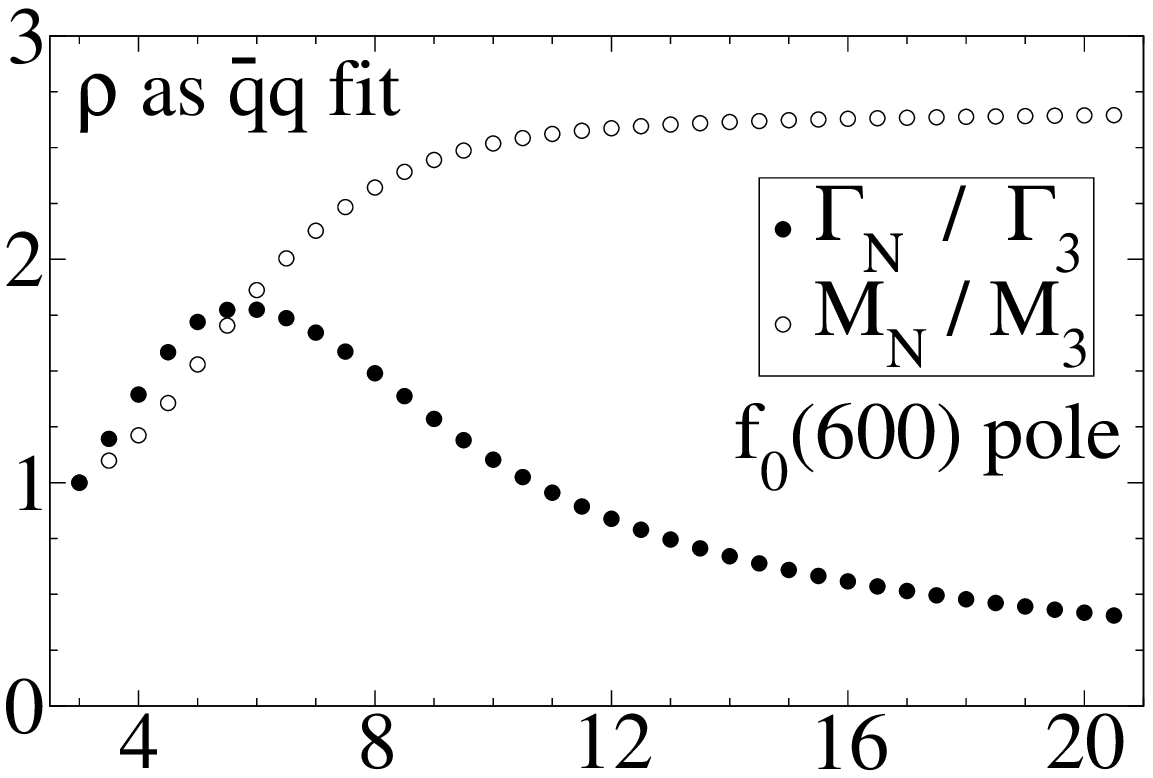}
    \includegraphics[angle=-90,scale=.34]{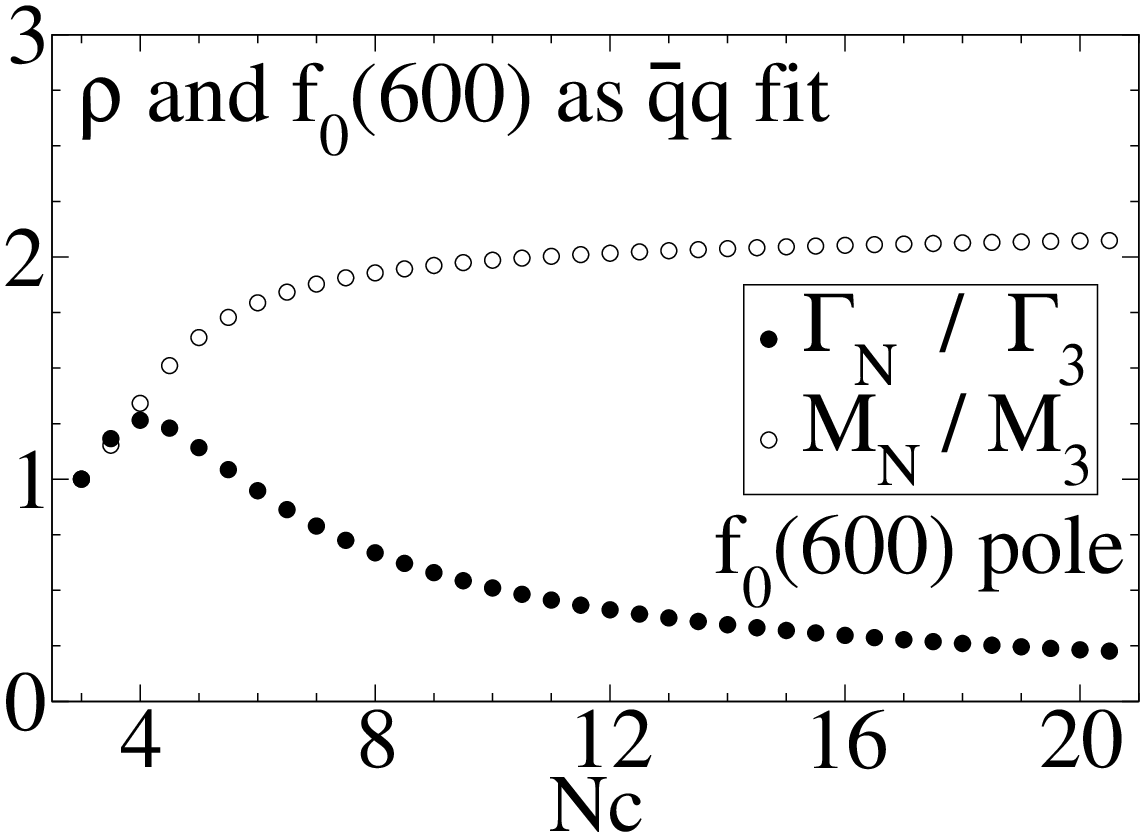}
  }
  \caption{{\bf Right and center}: $N_c$ behavior of the $\rho$
    and $\sigma$ pole at $O(p^6)$ with the ``$\rho$ as 
    $\bar qq$ fit''.
    {\bf Center:} Sigma behavior with $N_c$ at $O(p^6)$ with
    the ``$\sigma$ as $\bar qq$ fit''.}
  \label{2loops}
\end{figure}

\vspace*{-.4cm} 
\section{Discussion and conclusions}
We have seen that, within ChPT unitarized with the IAM,
the $N_c$ behavior of $\bar{q}q$ states is clearly identified 
whereas scalar mesons
behave differently near $N_c=3$. Here we want to emphasize again \cite{Pelaez:2005fd},
what can and {\it what cannot} be concluded from this behavior
and clarify some frequent questions and doubts raised in this meeting,
private discusions and the literature:
\newcounter{input}
\begin{list}{$\bullet$}{
\setlength{\leftmargin}{0.2cm}
\setlength{\labelsep}{0.cm}}

\vspace{-.2cm}
\item {\it The \underline{dominant component}  of the $\sigma$ and $\kappa$ {\it in meson-meson scattering} does not behave as a $\bar{q}q$}. Why ``dominant''?
    Because, most likely, scalars are a mixture of different kind of
    states.  If the $\bar{q}q$ was {\it dominant}, they would behave
    as the $\rho$ or the $K^*$ in Fig.1.  {\it But a smaller fraction of $\bar{q}q$ cannot be
      excluded.} Actually, it is somewhat favored in our $O(p^6)$ analysis \cite{Pelaez:2006nj}.
      
\vspace{-.2cm}
  \item {\it Two meson and some tetraquark states\cite{Jaffe} have a consistent
      ``qualitative'' behavior}, i.e., both disappear in
    the  meson-meson scattering continuum as $N_c$
    increases. Our results are not able yet
     to establish the nature of that dominant component.
The most we could state is that the behavior of
      two-meson states or some tetraquarks might be
qualitatively consistent.
\end{list}

The $N_c\rightarrow\infty$
limit has been studied in \cite{Sun:2004de,Nieves:2009ez}. Apart from its
mathematical interest, it could have some physical
relevance if the data and the large $N_c$ uncertainty on the
choice of scale were more accurate. Nevertheless:
\begin{list}{\roman{input}.}{\usecounter{input}
\setlength{\leftmargin}{0.2cm}
\setlength{\labelsep}{0.cm}}
\vspace{-.2cm}
\item[$\bullet$ ] As commented above, {\it a priori the IAM is not reliable in the  $N_c\rightarrow\infty$ limit}, since it corresponds to a weakly interacting theory, where 
exact unitarity
becomes less relevant in confront of other approximations made in the IAM derivation. It has been shown \cite{Nieves:2009ez} that it might work
well in that limit in the vector channel of QCD but not in the scalar channel.

\vspace{-.2cm}
\item[$\bullet$ ] Another reason to limit ourselves to $N_c$ not too
far from 3 is that in our calculations we have not included the $\eta'(980)$,
whose mass is related to the $U_A(1)$ anomaly and scales as 
$\sqrt{3/N_c}$.
Nevertheless, if in our calculations we keep $N_c<30$, its mass
would be $>310\,$MeV and thus pions are still the only relevant degrees
of freedom for the scalar channel in the $\sigma$ region.

\vspace{-.2cm}
\item[$\bullet$ ] 
{\it Contrary to the leading $1/N_c$ behavior 
\underline{in the vicinity of  $N_c=3$},
the $N_c\rightarrow\infty$ 
limit does not give information on the ``dominant component''
of light scalars.} The reason was 
commented above: In contrast to $\bar{q}q$
states, that become bound, 
 two-meson and some tetraquark
states dissolve in the 
continuum as $N_c\rightarrow\infty$. 
Thus, even if we started with an infinitesimal $\bar{q}q$ component
in a resonance, for a sufficiently large $N_c$ 
it may become dominant, and beyond 
that $N_c$ the associated pole
would behave as a $\bar{q}q$ state
although the original state only had an infinitesimal admixture of $\bar{q}q$.
Also, since the mixings of 
different components could change
with $N_c$, a too large $N_c$ could alter significantly
the original mixings. 
\end{list}

Actually, this is what happens for the one-loop IAM $\sigma$ resonance 
for $N_c\to\infty$, but 
it does {\it not} necessarily mean that 
the ``correct interpretation... is that
the $\sigma$ pole is a conventional 
$\bar{q}q$ meson environed by heavy pion clouds'' \cite{Sun:2004de}.
That the scalars are not conventional, is simply seen by comparing
them in Figs.1 and 2 with the ``conventional'' $\rho$ and $K^*$ in
those very same figures.
A large two-meson component is consistent,
but the $N_c\rightarrow\infty$ of the one-loop unitarized ChPT 
pole in the scalar channel
 limit is not unique \cite{Sun:2004de,Nieves:2009ez} 
given the uncertainty
in the chiral parameters. Moreover, for some LECS the scalar channel one-loop
IAM in the $N_c\rightarrow\infty$ limit can lead to phenomenological inconsistencies 
\cite{Nieves:2009ez} since poles can even move to negative 
mass square (weird), 
to infinity or to a  positive mass square. 
That is one of the reasons why in the figures here and in \cite{Pelaez:2003dy,Pelaez:2006nj}
we only plot up to 
$N_c=30$, but not 100, or a million.
Hence, robust 
conclusions on the dominant light scalar component,
can be obtained not too far
 from real life, say $N_c<15$ or 30,  for a $\mu$ choice between
roughly $0.5$ and 1 GeV, that simultaneously ensures the $\bar qq$ dependence for the $\rho$ and $K^*$ mesons.  Note, however, that under these same conditions the two-loop IAM still finds
a dominant non-$\bar qq$ component, but, in addition, a hint of a $\bar qq$ subdominant component, which
 is not conventional in the sense that it appears at a much higher mass than the physical $\sigma$.
This may support the existence of a second $\bar qq$ scalar octet above 1 GeV \cite{VanBeveren:1986ea}. 

{\it In summary, 
the dominant component of light scalars as 
generated from unitarized one loop ChPT 
scattering amplitudes does not behave as a
$\bar{q}q$ state as $N_c$ increases not far from $N_c=3$.
When using the two loop IAM
result in SU(2), below $N_c\sim\,$15 or 30,  
there is a hint of a subdominant $\bar{q}q$ component, but
 arising at roughly twice the mass of the physical $\sigma$.}

\end{document}